\documentclass[sigconf]{acmart}

\AtBeginDocument{}

\copyrightyear{2024}
\acmYear{2024}
\setcopyright{rightsretained}
\acmConference[CHI '24]{Proceedings of the CHI Conference on Human Factors in Computing Systems}{May 11--16, 2024}{Honolulu, HI, USA}
\acmBooktitle{Proceedings of the CHI Conference on Human Factors in Computing Systems (CHI '24), May 11--16, 2024, Honolulu, HI, USA}
\acmDOI{10.1145/3613904.3642243}
\acmISBN{979-8-4007-0330-0/24/05}

\usepackage{xspace}
\usepackage{calc}

\usepackage[resetlabels]{multibib}
\newcites{game}{Ludography}
\newcommand{\citegameprefix}{G}
\usepackage{xparse}
\let\origcitegame\citegame
\RenewDocumentCommand{\citegame}{O{} O{} m}{\renewcommand{\citenumfont}[1]{\citegameprefix##1}\origcitegame[#1][#2]{#3}\renewcommand{\citenumfont}[1]{##1}}

\begin{document}
\newcommand{\todo}[1]{}
\renewcommand{\todo}[1]{{\color{red} \textbf{TODO: {#1}}}}

\newcommand{\fix}[1]{}
\renewcommand{\fix}[1]{{\color{orange} \textbf{FIXME: {#1}}}}

\newcommand{\cit}[1]{}
\renewcommand{\cit}[1]{{\color{green} \textbf{CITEME: {#1}}}}

\newcommand{\system}{DungeonMaker\xspace}
\newcommand{\systemEmph}{\emph{\system}\xspace}
\newcommand{\systemEmphSp}{\systemEmph\xspace}

\newcommand{\iquote}[2]{\textit{``{#1}''}~({#2})}
\newcommand{\dirquote}[1]{\textit{``{#1}''}}

\renewcommand{\sectionautorefname}{Section}
\renewcommand{\subsectionautorefname}{Section}
\renewcommand{\subsubsectionautorefname}{Section}
\renewcommand*{\figureautorefname}{Fig.}

\title[\system]{\system: Embedding Tangible Creation and Destruction in Hybrid Board Games through Personal Fabrication Technology}

\author{Evgeny Stemasov}
\orcid{0000-0002-3748-6441}
\email{evgeny.stemasov@uni-ulm.de}
\author{Tobias Wagner}
\orcid{0000-0002-5907-5248}
\email{tobias.wagner@uni-ulm.de}
\affiliation{\institution{Ulm University}
  \city{Ulm}
  \country{Germany}
}

\author{Ali Askari}
\orcid{0000-0002-4374-3635}
\authornote{Both authors contributed equally to this research.}
\email{ali.askari@uni-ulm.de}
\author{Jessica Janek}
\orcid{0000-0002-7483-0096}
\authornotemark[1]
\email{jessica.janek@alumni.uni-ulm.de}
\affiliation{\institution{Ulm University}
  \city{Ulm}
  \country{Germany}
}

\author{Omid Rajabi}
\orcid{0000-0003-4498-2539}
\email{omid.rajabi@uni-ulm.de}
\author{Anja Schikorr}
\orcid{0000-0003-0842-5890}
\email{anja.schikorr@alumni.uni-ulm.de}
\affiliation{\institution{Ulm University}
  \city{Ulm}
  \country{Germany}
}

\author{Julian Frommel}
\orcid{0000-0001-8783-7783}
\email{j.frommel@uu.nl}
\affiliation{\institution{Utrecht University}
  \city{Utrecht}
  \country{Netherlands}
}

\author{Jan Gugenheimer}
\orcid{0000-0002-6466-3845}
\email{jan.gugenheimer@tu-darmstadt.de}
\affiliation{\institution{TU Darmstadt}
  \city{Darmstadt}
  \country{Germany}
}
\affiliation{\institution{Institut Polytechnique de Paris}
  \city{Paris}
  \country{France}
}

\author{Enrico Rukzio}
\orcid{0000-0002-4213-2226}
\email{enrico.rukzio@uni-ulm.de}
\affiliation{\institution{Ulm University}
  \city{Ulm}
  \country{Germany}
}

\renewcommand{\shortauthors}{Stemasov, et al.}

\begin{abstract}
  Hybrid board games (HBGs) augment their analog origins digitally (e.g., through apps) and are an increasingly popular pastime activity.
Continuous world and character development and customization, known to facilitate engagement in video games, remain rare in HBGs.
If present, they happen digitally or imaginarily, often leaving physical aspects generic.
We developed DungeonMaker, a fabrication-augmented HBG bridging physical and digital game elements: 1) the setup narrates a story and projects a digital game board onto a laser cutter; 2) DungeonMaker assesses player-crafted artifacts; 3) DungeonMaker's modified laser head senses and moves player- and non-player figures, and 4) can physically damage figures.
An evaluation ($n=4\times3$) indicated that DungeonMaker provides an engaging experience, may support players' connection to their figures, and potentially spark novices' interest in fabrication.
DungeonMaker provides a rich constellation to play HBGs by blending aspects of craft and automation to couple the physical and digital elements of an HBG tightly. \end{abstract}



\begin{teaserfigure}
\centering
  \includegraphics[width=\linewidth]{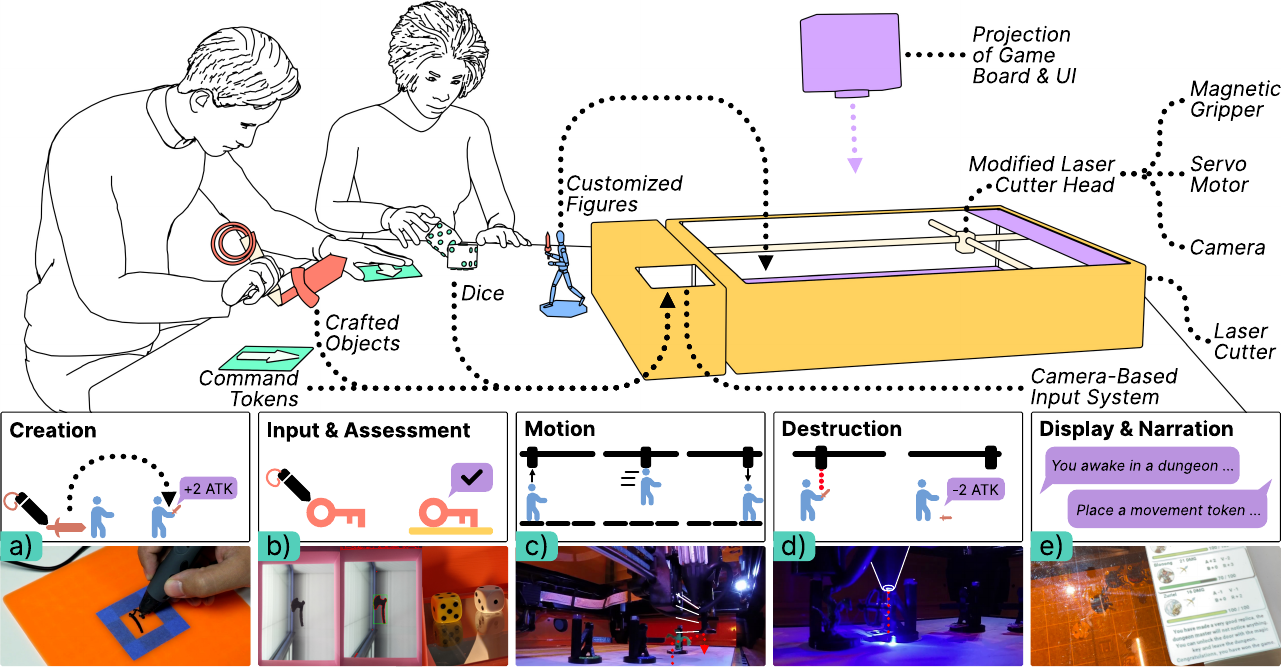}
  \caption{\system is a hybrid board game played using fabrication devices (laser cutter, 3D-printing pens) to tightly couple digital and physical game elements. a)~Players can create artifacts with a 3D-printing pen to personalize and improve their game figures; b)~\system can assess players' creations using a camera-based input system, which also processes dice- and token-based inputs; c)~a custom attachment to a laser head can grab and move figures according to player commands or game logic; d)~the laser cutter can also permanently damage figures; e)~\system unifies the aforementioned components into a tabletop role-playing experience while adding display and narration capabilities for an immersive experience.}
  \Description{}
  \label{fig:teaser}
\end{teaserfigure}

\maketitle

\section{Introduction}
\label{sec:introduction}
Board games have been a well-established pastime activity in many households \cite{rogersonLoveAllBits2016}.
They provide a playful, social, and tangible experience that usually engages multiple, co-located people in an informal setting~\cite{sousaBackGameModern2019,xuChoresAreFun2011}.
To further this analog concept, so-called \emph{hybrid} board games have emerged as a more recent phenomenon~\cite{cavicchiniHybridBoardGame2019, rogersonCapturingHybridityComparative2021, rogersonMoreGimmickDigital2021}.
Hybrid board games combine physical, tangible elements with digital elements (e.g., support tools) to augment the experience, for instance, by helping players learn rules, providing calculations, or storytelling elements \cite{rogersonUnpackingBoardgamesApps2021}.
In summary, hybrid board games aim to combine the tangible experience board games provide, with digital benefits, to \emph{augment} the player experience~\cite{jarvisDigitalLosingPhysical2022}.\\

While hybrid board games aim to augment the tangible, physical experience of traditional board games, they have little to no effect on the physical play space (e.g., players' figures or the game board). 
This constrains the design space for hybrid board games in different ways:
\textbf{(1)~Digital-physical synchronization}: State changes created via the digital part (e.g., a figure's position) have to be ``enacted'' by players to keep digital and physical representations in sync.
This constrains the influence of the digital part of a hybrid board game to simpler, digital-only aspects, like keeping track of statistics \cite{rogersonMoreGimmickDigital2021}.
In turn, this leaves an untapped potential for novel, hybrid board game experiences that connect and unify digital and physical elements.
\textbf{(2)~Absence of Personalization:} Unlike digital games, analog or hybrid board games allow for little to no \emph{personalization} of the physical artifacts involved: figures and boards, for instance, are used as-is, while changes, if any, are kept track of in the digital domain.
This limits the expressivity of the game, while also hindering players' attachment to their representations (i.e., avatar/figure) during the game, as they are left with generic physical representations of themselves.
Personalization is often manifested in board games using character sheets, most commonly known from games like Dungeons and Dragons~\citegame{dnd}, but also present in legacy-style games like Pandemic Legacy~\citegame{pandemiclegacyseason1}, which are missing a digital dimension and rarely manifest in physical space beyond paper elements.
Such connection with characters and avatars, however, is an essential component of numerous \emph{digital} game experiences~\cite{birkFosteringIntrinsicMotivation2016, kaoEffectsBadgesAvatar2018, turkayEffectsAvatarbasedCustomization2015}, highlighting the potential for fostering such an experience in hybrid board games.
\textbf{(3)~Absence of Physical Risk and Continuity:} Lastly, there are few persistent, lasting effects possible and feasible in hybrid board games: damage, destruction, and loss happen in the digital domain and are often non-permanent.
An outlook of permanent loss, however, can be beneficial for player experiences (e.g., cf. permadeath in DayZ~\citegame{dayz}\cite{allisonGoodFrustrationsParadoxical2015,carterFearLossMeaningful2017}), and support notions of risk and consequences~\cite{carterFearLossMeaningful2017} when used as a conscious design element~\cite{rossmyPointNoUndo2023,rossmyPunishableAIExamining2020}. 
Permadeath is indeed present in several non-hybrid board games like Pandemic Legacy~\citegame{pandemiclegacyseason1}, where it remains in the physical domain and has to be enacted by the players.
Such experiences could be fostered by integrating \emph{lasting} and \emph{physical} effects into hybrid board games.\\

To introduce and probe these dimensions currently largely absent from or underexplored in hybrid board games, we developed \system.
\system is a hybrid board game, collaboratively played by 3 players using a set of craft and fabrication tools (\autoref{fig:teaser}).
It is a tabletop RPG (Role-Playing Game), presenting a scenario where 3 players are trapped in a dungeon and must coordinate their escape.
Each player controls one figure (representing their avatar), which they can customize with a 3D printing pen to their liking.
The players have to gather clues about how to escape a dungeon, while also engaging with enemies spawning on the board.
During the game, users get the opportunity to fabricate weapons for their figures with a 3D-printing pen (\autoref{fig:teaser}a), which are evaluated by the system and given gameplay-relevant stats (e.g., damage, \autoref{fig:teaser}b).
The figure customization and weapon-making introduce an aspect of \emph{physical creation} and \emph{personalization} to the game, as users craft and attach these artifacts to their own figure, which serves as their avatar in the game.
Figure movement is executed by the laser cutter (\autoref{fig:teaser}c) following user input through tangible tokens.
In the case of a defeat during a battle, the laser cutter may damage a figure, along with the weapon made by the player (\autoref{fig:teaser}d).
This introduces permanent, physical \emph{destruction} to the gameplay.
The experience is further supported through audio narration and a game board projected onto and into the laser cutter (\autoref{fig:teaser}e).
In sum, \system embeds both additive and subtractive manufacturing approaches but deliberately does not do so with the goal of solving engineering or design challenges, but rather to augment hybrid board games, further unifying their digital and physical representations through playful interaction with and through fabrication devices.\\

We probed how users engage with \system in an exploratory evaluation to understand how fabrication-augmented hybrid board games may alter the player experience and connection to players' figures.
Groups of 3 were invited for game sessions taking up to 50 minutes.
This was the approximate duration of one scenario, including all (deliberately time-constrained) fabrication activities.
Through our evaluation, we found that \system provides an engaging and enjoyable experience, increases players' perceived connection to their figures, and may even serve as a point of first contact with personal fabrication, as it increased self-reported curiosity for fabrication in novices. 
Based on the evaluation and \system's development, we reflect on design considerations relevant for fabrication-augmented hybrid board games. We argue that \system is a novel instance of infusing current hybrid board games with fabrication concepts to enable a tighter coupling between the physical and digital environment. 
This leads to a stronger relationship between players and game experience through personalization and the potential for tangible loss.\\

The contributions of our work are the following:
\begin{enumerate}
    \item Development of \system, a collaborative hybrid board game played using a set of fabrication devices. The setup enables the creation and alteration (i.e., personalization), destruction, and movement of physical figures and their attached equipment, processes commands based on tangible inputs, and uses projection to display a dynamic game board, along with the game state.
    \item Insights gathered from an evaluation of \system with 4 triads ($n = 12 = 4 \times 3 $), indicating an engaging, enjoyable player experience and players' increased identification with their figures after a game session. \item Design considerations for fabrication-augmented hybrid board games through the lenses of technology accessibility, value creation, and personalization.
\end{enumerate} \section{Related Work}
\label{sec:relatedwork}
    \system is inspired by prior works in the domains of (hybrid) board games, playful craft, and fabrication approaches, personalization in games, and augmented or unconventional uses of fabrication devices.
    It is further inspired by established commercial board games (e.g. legacy games~\citegame{risklegacy,pandemiclegacyseason1}, wargames~\citegame{warhammerthirdedition,fullthrust}, hybrid board games~\citegame{xcomtheboardgame}, and games with craft components~\citegame{exitthegamethesecretlab}), whose influence we outline in \autoref{sec:designprocess}.
    
    \subsection{Hybrid Board Games}    
       Hybrid board games are defined as combinations of physical board games and digital artifacts ``uniting''~\cite{rogersonPrecursorsModernHybrid2020}, ``augmenting''~\cite{tanenbaumDesigningHybridGames2017}, or ``blending''~\cite{arjorantaBlendingHybridGames2016} digital and physical elements.
       We situate \system as a step beyond disjoint digital and physical game spaces in hybrid games, by coupling them further through creation and destruction processes.
       Rogerson et al. identified roles of digital tools in hybrid board games (e.g., calculation, storytelling, randomization)~\cite{rogersonUnpackingBoardgamesApps2021}, attitudes towards digital augmentations \cite{rogersonMoreGimmickDigital2021}, and analyzed existing commercial hybrid board games \cite{rogersonCapturingHybridityComparative2021}.
        Mandryk et al. \cite{mandrykFalseProphetsExploring2002} and Kankainen et al. \cite{kankainenInterplayTwoWorlds2016} further explored social components \cite{nummenmaaSocialFeaturesHybrid2019} of hybrid board games, as they mediate social interactions in addition to interactions with the game itself \cite{grasseMadMixologistExploring2021}. 
        More specific design guidelines are also present in the space, and, for instance, include ``customizability'' \cite{kankainenHybridBoardGame2019} or ``aesthetics''~\cite{kankainenHybridBoardGame2019}, aspects we embrace with \system.
        Kankainen and Paavilainen further mention ``tangibility''~\cite{kankainenHybridBoardGame2019} and ``tutorials''~\cite{kankainenHybridBoardGame2019} as relevant guidelines, which we similarly enable and support with our implementation.
        Narration-focused experiences were also explored for analog board games \cite{sullivanTaxonomyNarrativecentricBoard2017} and hybrid board games \cite{obrienPantheonDreamPlaying2017}, highlighting the aspect of immersion.
        Approaching hybrid board games from a technical perspective, G??mez-Maureira highlighted the relevance of autonomous components in hybrid board games and the potential of artificial intelligence \cite{gomez-maureiraTaxonomyAIHybrid2020}.
        Similarly, Jensen et al. augmented ``Settlers of Catan''~\citegame{thesettlersofcatan} using electrochromic displays that can dynamically change~\cite{jensenHybridSettlersIntegrating2020}, which embodies an alternative approach to app-based hybridization~\cite{rogersonUnpackingBoardgamesApps2021}.
        
        Despite their hybrid nature (i.e., being situated in physical and digital spaces simultaneously), hybrid board games only control the digital game space, with no direct control over the physical one.
        \system leverages fabrication devices to allow the game control over this physical space (both through manipulations like motion, and through destruction), while also allowing players to contribute to the physical play space in a creative and functional (i.e., gameplay-relevant) fashion.
        
    \subsection{Craft and Fabrication Games}
        Expression through craft and design as a gameplay element has also been considered in game experiences.
        With the emergence of accessible digital fabrication devices~\cite{baudischPersonalFabrication2017, motaRisePersonalFabrication2011}, this trend was further reinforced.
        Sullivan and Smith specified aspects and categories of ``craft games''~\cite{sullivanDesigningCraftGames2016}.
        \system falls under two categories of their classification, as it is a game  
        \dirquote{[...] in which crafting is part of the core mechanics}~\cite{sullivanDesigningCraftGames2016} but also one where \dirquote{crafting machines or tools [are] used as game interfaces}~\cite{sullivanDesigningCraftGames2016}.
        Tanenbaum et al. introduced the notion of ``playful fabrication'', arguing for uses of fabrication technology beyond prototyping \cite{tanenbaumDesigningHybridGames2017}, a concept we embrace with \system, along with the patterns of idleness, augmentation, and accumulation presented in the work \cite{tanenbaumDesigningHybridGames2017}.
        ``Fabrication Games'' by Bhaduri et al. similarly outlines several ideas, like the creation of game pieces, mementos, or input capture \cite{bhaduriFabricationGamesUsing2017}.
        By adding, destruction, automated motion, narration, and visual output, we continue this line of thought and augment the hybrid board game experience through fabrication devices.
        
        ``Exquisite Circuits'' by Qi et al. used the game of ``exquisite corpse'' to enable the collaborative and playful creation of electronic circuits~\cite{qiExquisiteCircuitsCollaborative2021}.
        Hybrid Embroidery Games by Lee and Albaugh presented textile-based game systems, similarly enabling playful interaction with fabrication devices, while further highlighting social experiences \cite{leeHybridEmbroideryGames2021}.
        Grasse and Melcer developed ``Generation'', a game that requires players to fabricate and assemble components to progress~\cite{grasseGenerationNovelFabrication2020}. 
        Notably, the elements are 3D-printed during the game, and they are digitized using a tablet~\cite{grasseGenerationNovelFabrication2020}, creating a constellation resembling a hybrid board game.
        ``Threadsteading'' by Albaugh et al. presented a hybrid board game played using an embroidery machine \cite{albaughThreadsteadingPlayfulInteraction2016}.
        Notably, each game session yields a quilt that is comprised of traces of the game session--a tangible memento.
        This is comparable to the work of Sullivan et al., who used a Loom to interact with a game while fabricating a tangible memento of the game \cite{sullivanLoominaryCraftingTangible2018}.
        Eickhoff et al. presented ``Destructive Games'' using a laser cutter \cite{eickhoffDestructiveGamesCreating2016}.
        With \system, we explore applications beyond minigames, but rather more complex scenarios, and we further allow users to create artifacts that meaningfully influence gameplay and can be lost during it. 
        This is also related to tabletop roleplaying games that similarly sometimes feature the destruction of components~\cite{engelsteinBuildingBlocksTabletop2022}. Such games have already been explored in the context of hybridity, for example, by Buruk et al.~\cite{burukAugmentedTableTopRolePlaying2017}, who developed interactive digital artifacts to enable and explore augmented tabletop RPGs. 
        ``FabO'' presented a novel workflow to add fabrication activities to existing digital games, for instance, to create mementos of a salient moment in a game \cite{turakhiaFabOIntegratingFabrication2021}.
        This approach was further evaluated with a focus on aesthetics \cite{turakhiaIdentifyingGameMechanics2022}. 
        Embedding craft and (digital) fabrication in game experiences augments them by providing them with ways to control and influence physical spaces, not only digital ones. 
        The desire of players to use fabrication in board game play is also evident in recent work. \citet{tchernavskijReadymadesRepertoiresArtifactMediated2022} found that game masters in tabletop roleplaying games made use of devices like paper printers and laminating machines when preparing for their sessions. Such approaches were also used by other board game players. In their study on homebrewed hybridity during COVID-19 times, \citet{sparrowLessonsHomebrewedHybridity2023} found that distanced boardgame players used at-home fabrication devices such as printers and 3D printers. 
        
        With \system, we embrace these notions and apply them to hybrid board games, facilitating not only creation and personalization but also the potential for physical damage and loss.

    \subsection{Personalization and Agency in Games}
        Games are influenced by a multitude of constructs that define the player experience as a whole. 
Digital games provide ways to personalize one's avatar -- a representation perceived by others, but also one that players themselves perceive~\cite{sibillaAmNotMy2018}.
        Turkay and Adinolf determined a positive effect of customization on motivation \cite{turkayEffectsCustomizationMotivation2015}.
        This was further confirmed by Birk et al., who emphasize intrinsic motivation as a core factor \cite{birkFosteringIntrinsicMotivation2016}.
        Kao and Harrell explored the effects of badges as a personalization element \cite{kaoEffectsBadgesAvatar2018}.
        In the context of immersive environments (Virtual Reality, VR), Cuthbert et al. confirmed that avatar customization has a positive effect on players  \cite{cuthbertEffectsCustomisationPlayer2020}, which was echoed by Kouloris et al. in the context of VR exergames \cite{koulourisMeVsSuper2020}.
        While there is an aspect of personalization and self-expression in the context of Warhammer \cite{harropEveryoneWinnerWarhammer2013} tabletop games, for instance, they do not rely on machine-driven influences, and assessments and usually involve no lasting physical consequences \cite{carterDraftingArmyPlayful2014}.
        
        With \system, we also alter notions of \emph{agency} in board games, by re-assigning roles (e.g., moving figures).
        Abeele et al., for instance, emphasized how curiosity and autonomy are crucial components of player experience \cite{abeeleDevelopmentValidationPlayer2020}. 
        Thue et al. highlighted how agency in the context of games is highly relevant and positively influences the experience, making interactive applications such as games powerful candidates to hand agency to users \cite{thuePlayerAgencyRelevance2010} and foster it through perceptible control \cite{thompsonIllusionsControlUnderestimations1998,seinfeldUserRepresentationsHumanComputer2021}.

        In games research, the concept of permadeath (i.e., characters becoming permanently unavailable) is also present in both digital games and board games.
        West et al. state that a prospect of permanent loss may create meaning \cite{westItAllFun2022}.
        This is echoed by Carter and Allison for the multiplayer shooter ``DayZ''~\citegame{dayz}, highlighting how negative experiences may intensify the experience~\cite{carterFearLossMeaningful2017}.

        Personalization and attachment are highly relevant factors in digital games, but are largely absent from physical components of analog board games (apart from role-playing games; \citegame{dnd,kingdomdeathmonster}) and from hybrid board games.
        This constrains the design space of such games, despite their inherent materiality \cite{rogersonLoveAllBits2016}.
        With digital fabrication and craft technologies, personalization can be made possible for hybrid board games, by handing players ways to customize their avatars to increase identification and handing games ways to alter and destroy artifacts players may have grown attached to, echoing notions of loss and ``permadeath''~\cite{carterFearLossMeaningful2017,westItAllFun2022} in physical game experiences.
        
    \subsection{Expanded Uses of Fabrication Devices}
        \system leverages available fabrication devices to enable new behaviors and applications.
        In particular, it relies on computer numerical control (CNC), which allows for motion control, commonly applied to manufacturing.
        This notion has been expanded upon by Katakura et al., who augmented a commercial 3D-printer to be able to assemble objects or remove support material \cite{katakura3DPrinterHead2019}.
        Teibrich et al. added a mill to a 3D-printer head to support the notion of ``patching'' physical objects instead of re-fabricating them~\cite{teibrichPatchingPhysicalObjects2015}.
        LaserFactory by Nisser et al. leveraged a laser cutter to fabricate functional objects within the device's workspace~\cite{nisserLaserFactoryLaserCutterbased2021}, essentially expanding the device's output capabilities.
        Augmented reality (AR) has also been applied to augment design tools~\cite{stemasovMixMatchOmitting2020,stemasovShapeFindARExploringInSitu2022,weichelMixFabMixedrealityEnvironment2014} or expand the output capabilities of fabrication devices.
        Ludwig et al. used this to allow meaningful, in-situ output from a 3D-printer \cite{ludwigPrinterTellingMe2019}.
        Olwal et al. \cite{olwalSpatialAugmentedReality2008} applied this notion to industrial CNC machinery.

        Notions of unmaking \cite{songUnmakingEnablingCelebrating2021,wuUnfabricateDesigningSmart2020}, ephemerality \cite{stemasovEphemeralFabricationExploring2022}, and destruction \cite{muellerScottyRelocatingPhysical2015} in the context of digital fabrication have also been explored as a design material and interaction paradigm.
        Role assignments between humans and fabrication devices have also been questioned in the works of Devendorf et al.~\cite{devendorfBeingMachineReconfiguring2015, devendorfProbingPotentialPostAnthropocentric2016, devendorfReimaginingDigitalFabrication2015}.
        
        \system is inspired by these works and similarly reconfigures how audiences engage with fabrication devices to enhance the experience of hybrid board games.
        We leverage CNC as a foundation~\cite{katakura3DPrinterHead2019} to allow a laser cutter to translate figures, and enrich this constellation through tangible~\cite{stemasovBrickStARtEnablingInsitu2023}, self-made inputs and dynamic in-situ outputs (i.e., projection and audio). \section{Implementation of \system}\label{sec:system}
      \begin{figure*}[h!]
          \centering
          \includegraphics[width=\linewidth]{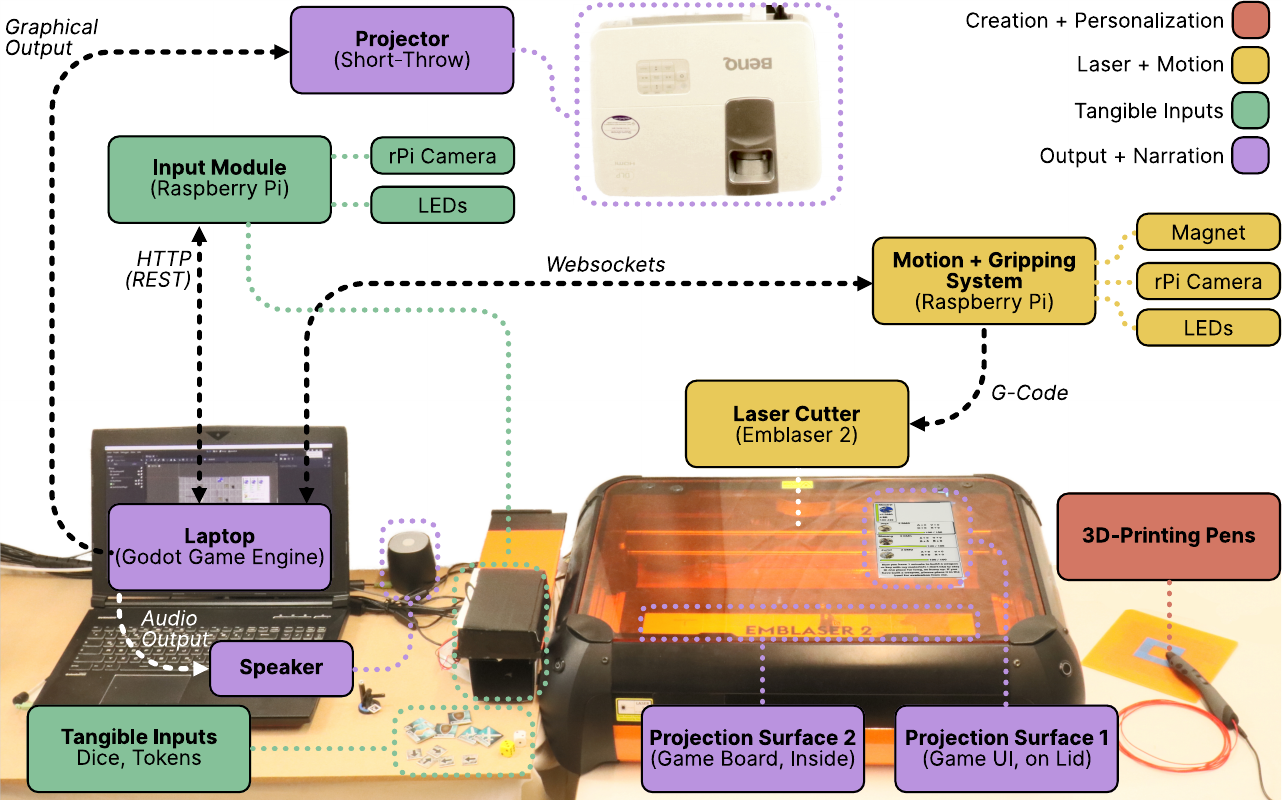}
          \caption{Overview of all components used in \system and how they are interconnected. The core components are the \system game, the camera-based input system (used to scan tokens, dice, or crafted inputs), and the motion control system connected to the laser cutter. Visual output is handled through a projector, rendering onto two surfaces: the lid of the laser cutter and its working area (the top projection in this figure was composited for clarity).}\label{fig:hw-overview}
          \Description{}
        \end{figure*}
        
    The following sections first outline the design rationale we followed for the development of \system.
    This is followed by a description of the technical setup and the game we implemented.
    
    \subsection{Design Rationale}
        With both the hardware setup of \system, and the implemented game, we aimed to cover an extensive range of design components within the space of combining hybrid board games and (personal) fabrication technologies.
        To do so, we initially outlined a design space, which combined fundamental actions in (hybrid) board games \cite{rogersonMoreGimmickDigital2021,rogersonUnpackingBoardgamesApps2021,kankainenHybridBoardGame2019} (e.g., motion, calculations, narration) and fundamental actions enabled by fabrication devices \cite{tanenbaumDesigningHybridGames2017,turakhiaIdentifyingGameMechanics2022,baudischPersonalFabrication2017} (e.g., motion, material addition, material removal).
        Each resulting cell can then be used to generate (inter)actions between players and machines. 
        For instance, using additive manufacturing for narration may require a pen plotter or 3D printer to print text onto the game board.
        These types of alterations, re-assignments (i.e., changes to role distributions between players and machines), or augmentations (i.e., additions to established board game experiences, like manufacturing) were then explored in a formative survey, which is described next.

        \subsubsection{Formative Survey}
            \label{sec:formativesurvey}
    We initially conducted a formative survey ($n=68$) to probe stances towards alternative role distributions between players and digital systems in (hybrid) board games, along with prospects of physical loss and damage. 
    
    \paragraph{Sample:} 
        The participants were recruited across our personal networks. 
        Filling out the survey took participants around 17 minutes, on average.
        Their age ranged from 20 to 59 ($M = 28.8, SD = 8.6$); 31 of them were students, 29 were employed, with the remaining occupations being self-employed, apprentices, or homemakers.
        41 participants identified as male, 24 identified as female, and three participants preferred not to disclose their gender.
30 participants reported having played board games 1-6 times in the last 6 months, 19 participants played 6-12 times, and 13 played over 12 times.
        For 27 participants, the ideal duration for a board game session was 45-60 minutes. 
        15 participants preferred durations between 60 and 90 minutes, and only 3 participants preferred game session durations of over 90 minutes.
        32 out of 68 participants have played multi-session games (e.g., ``Dungeons and Dragons''~\citegame{dnd}). 
        When it came to \textbf{purely digital board games}, 31 out of our participants reported having played a game with some degree of digital assistance (e.g., ``Catan Classic''~\citegame{thesettlersofcatan}).
29 participants reported that they had played \textbf{hybrid board games} before (e.g., ``XCOM the Board Game''~\citegame{xcomtheboardgame}).

    \paragraph{Re-Assigning Roles: } 
We specifically wanted to understand sentiments towards (potentially unconventional) changes to role distributions in hybrid board games, where a device may be responsible for more than calculation support or narration.
        Participants were asked to rate a set of statements regarding ``leaving actions in a board game to a technical device'' on a 6-point Likert scale.
        
Tendencies to leave specific tasks to a digital system were found in tasks like 
        the motion of non-player game pieces (69\% agree),
        explaining rules (79\% agree),
        distance measurements (62\% agree),
        calculating cores (79\% agree),
        or serving as a narrator or game ``master'' (68\% agree).
        Additionally, 59\% of participants could see a system creating and modifying game pieces.
Participants had differing opinions regarding tasks like 
        moving figures assigned to a player (46\% agree) and
        destroying figures assigned to another player/opponent (51\% agree).
Lastly, the survey participants had reservations about leaving certain tasks to a digital system.
        This included 
        rolling dice (40\% agree), 
        destroying figures assigned to the player (44\% agree), or
        displaying possible moves (44\% agree).
        These results echo sentiments present in literature \cite{rogersonLoveAllBits2016,rogersonDigitisingBoardgamesIssues2015}, where users value tangibility and agency while being open to relinquishing certain tasks like scorekeeping to digital systems. 
    
    \paragraph{Loss and Destruction}
Lastly, we wanted to get an initial understanding of how potential users may react to the prospect of physical loss in hybrid board games.
        Permanent destruction has been presented as a design element in HCI literature~\cite{rossmyPointNoUndo2023}.
        Rossmy et al. outlined how physical alteration can be value-adding or value-removing and turns thoughtless interactions into thoughtful ones, if they are known to be irreversible~\cite{rossmyPointNoUndo2023}.
        With \system, we aimed to embrace these notions in the context of hybrid board games and the coupling of their physical and digital components.
        Participants were asked to imagine a scenario where they were bringing their own figures to a board game session.
        In the described scenario, it would be possible for the figures to be permanently damaged or destroyed. 
For self-made (e.g., 3D-printed) figures, 46\% reported agreement that they would be willing to bring them to such a game session.
        In contrast, purchased game pieces were less likely to be brought into such an environment, with only 31\% of participants reporting agreement here.\\
        
While our sample is likely biased toward board game enthusiasts (evident in their experience of more complex~\cite{carterDraftingArmyPlayful2014} board games, such as ``Warhammer''~\citegame{warhammerthirdedition}), the survey helped us understand sentiments towards re-assigned roles and the absence of tangible loss in hybrid board games.
In particular, users are willing to relinquish some of the roles they may be used to in board games and may be willing to engage with destruction in the context of board games, depending on their investment.
        To explore notions of craft and value-creation with a prospect of damage, we chose to embed these components in \system, but circumvent \emph{financial value} and loss \cite{eickhoffDestructiveGamesCreating2016,harropEveryoneWinnerWarhammer2013}. 
        Similarly, while we remove  ``chores''~\cite{xuChoresAreFun2011,rogersonDigitisingBoardgamesIssues2015} like movement or scorekeeping, we introduce new ones, like crafting.

        \subsubsection{Design Goals}\label{sec:designgoals}
            Based on the formative survey, along with practical and theoretical considerations found in the literature, we outlined the following high-level design goals for the development of \system:
            \begin{enumerate}
                \item[\textbf{DG1}] \textbf{Personalization:} Provide ways to express avatar personalization in a tangible fashion, replicating identification-increasing effects known from digital games and analog RPGs. Additionally, personalization may have not only aesthetic but also gameplay-relevant effects -- \cite{whitePlayerCharacterWhatYou2014,turkayEffectsAvatarbasedCustomization2015}.
                \item[\textbf{DG2}] \textbf{Physicality and Continuity:} Enable tangible, physical consequences of decisions that may transcend a single game session -- \cite{rossmyPointNoUndo2023,eickhoffDestructiveGamesCreating2016,engelsteinBuildingBlocksTabletop2022}.
                \item[\textbf{DG3}] \textbf{Bidirectionality between Digital and Physical:} Provide ways for the digital part of a hybrid board game to enact computational decisions on physical game elements (e.g., game board, figures) -- \cite{kaimotoSketchedRealitySketching2022,zhuMechARspaceAuthoringSystem2022,weichelReFormIntegratingPhysical2015}.
            \end{enumerate}

        \subsubsection{Game Design Process}\label{sec:designprocess}
            The following paragraphs briefly outline elements of the design and playtesting process used for \system.
            Initial inspirations were established games like Full Thrust~\citegame{fullthrust}, EXIT~\citegame{exitthegamethesecretlab}, Monopoly~\citegame{monopoly}, Dungeon Fighter~\citegame{dungeonfightersecondedition} and Stuffed Fables~\citegame{stuffedfables}, which all are not only complex, but also have complex individual components to them (e.g., non-trivial motion and damage calculation in Full Thrust~\citegame{fullthrust}).
            We analyzed the aforementioned games with respect to components such as lasting changes, time pressure, opportunities for creative expression (functional and aesthetic), changes to the map or environment, story, motion, goals, constellations (player-versus-player or player-versus-environment), and resources (e.g., currencies).  
            As an initial design exercise, all aforementioned components were considered through the lens of fabrication devices providing aspects like motion, creation, or damage.
            
            \paragraph{Gameplay} 
            Initial development of the game and its rules was done using a collaborative online whiteboard tool (Miro\footnote{\url{https://miro.com/}, Accessed: 29.01.2024}), akin to a paper prototype.
            Here, a digital variant of a board, was created, and 3 of the authors played sessions on the whiteboard using dice simulators.
            This helped choose the game flow, find initial statlines for equipment and NPCs (non-player characters), or discard cumbersome concepts like dedicated noise-dice\footnote{The noise concept remained in the game, but is calculated implicitly from the movement decisions (proximity to NPCs and travelled distance) players make}.
            Aspects like the size of the game grid were informed by two aspects: scenario duration and hardware capabilities.
            This highlights the cross-pollination between individual elements of \system across digital and physical domains.
            Specifically, the figure size was informed by the hardware constraints (e.g., for reliable gripping).
            The figure size, in turn, informed the grid size of the final game prototype.
            After the prototype, consisting of the digital and physical game elements, was playable, several testing sessions for balancing followed, also done by 3 of the authors.
            
            \paragraph{Assessment of Player-Crafted Objects} 
            Given that equipment fabricated by players can be highly diverse, especially when they have not had any experience using a 3D printing pen, a formative evaluation of crafted artifacts and their ratings was done.
            9 participants were asked to use the 3D printing pen to create an axe, a sword, and a bow, \system's basic weapon types. 
            All participants had no prior exposure to a 3D-printing pen.
            This exploration was done informally by one of the authors to understand how novices may use a 3D-printing pen and to inform thresholds of the recognition algorithm of \system.
            The resulting ratings were used to define the rarity (i.e., power) levels that \system derives from a contour similarity value (cf. \autoref{sec:crafting}). 
            
            \paragraph{Fine-tuning} Prior to the evaluation (\autoref{sec:evaluation}), more testing and fine-tuning (e.g., of random effects) was done by 2 of the authors to optimize the game for a study session of 30-45 minutes, to accommodate pre- and post-study questionnaires in a reasonable timeframe.
            We do not argue that the resulting gameplay and balance are impeccable (cf. \autoref{sec:evaluation}, where potential for improvement became apparent), but consider that they enabled an enjoyable experience, with the relevant values (e.g., weapon power, enemy counts, enemy statlines) being ``in the ballpark'' and suitable for a short and enjoyable session.

    \subsection{Technical Setup}\label{sec:setup}
        The following sections focus on the hardware constellation that enables \system as a platform for fabrication-augmented hybrid board game experiences.
        An overview of the components can be seen in \autoref{fig:hw-overview}.
        The core components are a laptop running the game using the Godot game engine, a laser cutter (Emblaser 2 by Darkly Labs\footnote{\url{https://darklylabs.com/emblaser2/}, Accessed: 28.01.2024}), and two Raspberry Pi\footnote{\url{https://www.raspberrypi.com/products/raspberry-pi-4-model-b/}, \mbox{Accessed:}~28.01.2024} 4 single-board computers -- one connected to the laser cutter and the gripper, and responsible for motion control and gripping, the other connected to a camera below the input area and responsible for parsing inputs (dice, tokens, crafted weapons) placed there.
        Additionally, we designed a custom attachment to the laser cutter nozzle to enable sensing, gripping, and re-positioning of players' figures on the game board.
        The laptop running the game serves as a central point for communications and is connected to the Input Module using HTTP/REST, and to the Motion and Gripping System using websockets. 
        The projector is connected to the laptop and renders the game onto/into the laser cutter while also outputting sound (i.e., narration) through a speaker.
    
        \subsubsection{Motion Control and Gripper}
            \begin{figure}[h!]
              \centering
              \includegraphics[width=\minof{\columnwidth}{0.6\textwidth}]{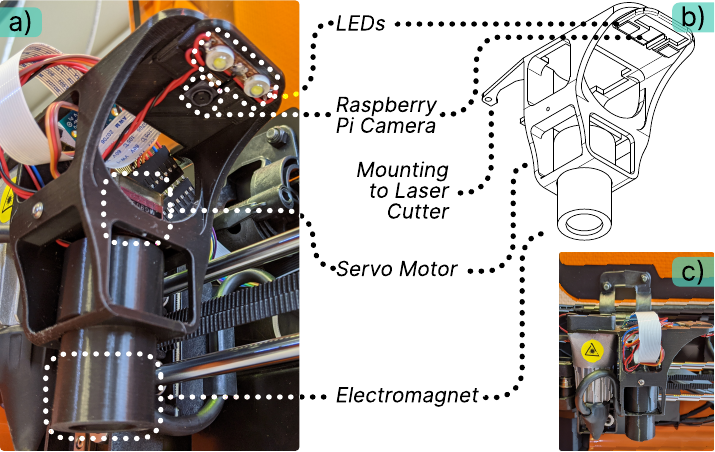}
              \caption{\system adds a module to the laser head~(a), consisting of an electromagnet for gripping figures, a servo motor to rotate them, a camera module~(a/b, top) to recognize figures on the board, supported by LEDs to ensure a well-lit workspace. The attachment design~(c) allowed the laser cutter to remain operational as a fabrication device despite the modification.}\label{fig:grabber}
              \Description{}
            \end{figure}
            
            The control software to enable motion and grabbing of figures uses an adapted version of Pronterface\footnote{\url{https://github.com/kliment/Printrun}, Accessed: 28.01.2024}.
            A Raspberry Pi 4 runs the software and connects to the laser cutter via USB.
            A custom 3D-printed attachment to the laser head (\autoref{fig:grabber}) enables \system to sense and move figures on the game board.
            In addition to serving motion control commands using G-code, the software can toggle the electromagnet at the gripper, and control the servo motor that can rotate the magnet.
            The electromagnet is toggled through the software via GPIO (General Purpose Input/Output) pins of the Raspberry Pi and an NPN transistor, but receives power from an external 5V power supply.

            \paragraph{Figure Movement}
            To grab a figure from the board, the modified laser head has to be able to detect the figures. 
            To do so, a camera, supported by 2 LEDs, is pointed downwards to recognize ArUco~\cite{garrido-juradoAutomaticGenerationDetection2014} markers attached to the base of each figure. 
            To re-position a figure, the head is first moved to the grid cell the figure is supposed to be in (as per game logic).
            Then, the camera detects the position and orientation of the marker, calculating the offset needed to be able to grab the figure correctly.
            This offset is then applied to the laser head, which executes a correction movement to position the magnet above the metal part of the figure.
            The laser head then lowers the z-axis, turns on the magnet, raises the z-axis again to clear other figures, and moves the figure to the target cell, releasing it there.
            An overview of the modified laser head is seen in \autoref{fig:grabber}.
            Notably, the laser cutter remains fully functional and usable as a fabrication device, despite these modifications, apart from a minor reduction in the work area in the x dimension (reduced from 50cm to approximately 40cm). 
            
            \paragraph{Figure Design}
            \system requires a special figure design to enable tracking and motion.
            Figures are mounted on a 3D-printed base (labeled as base plate), which provides an attachment for an ArUco marker and provides an interface used by the gripping mechanism (cf. \autoref{fig:physical-parts}a)-b)).
            Given that the gripper relies on a magnet, this interface consists of a metal nut attached to a plastic rod protruding from the base.
            They are also standardized with respect to their height, to ensure that the laser is focused when damaging figures. 
            The arms are also offset from the base, to ensure that the laser cutter, or melting plastic, do not damage the ArUco marker.
            Our current figure design is \emph{modular} and allows for the replacement of damaged arms rather easily.
            While this approach was applied to the player figures used in the study, there are few limitations beyond the base design and size constraints.
            The dragon NPC (cf. \autoref{fig:physical-parts}c), for instance, was handcrafted using a 3D-printing pen, combining the design with the more precisely manufactured base~\cite{takahashi3DPen3D2019}.
            Similarly, the generic design is meant to leave room for personalization. 
        
        \subsubsection{Tangible Inputs}
            \begin{figure}[h!]
              \centering
              \includegraphics[width=\minof{\columnwidth}{0.6\textwidth}]{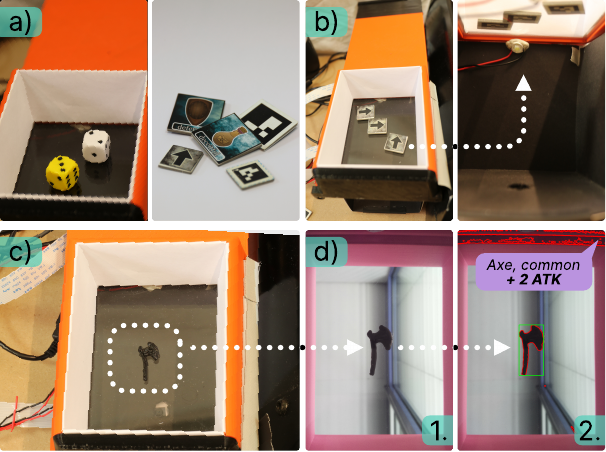}
              \caption{Tangible inputs handled by \system: a) dice are recognized using blob detection, while tokens have a fiducial marker on their backside. b) to move a figure, players place a set of motion tokens, which are converted to a path (here: right--right--up); c) crafted weapons are put on the plate, and contours are extracted d) and matched with a reference shape. The similarity score defines the game-relevant value of the crafted artifact.}\label{fig:tangible-inputs}
              \Description{}
            \end{figure}
            
            \system is able to process a set of tangible inputs made by players using the Input Module (\autoref{fig:tangible-inputs}). 
            Specifically, this enables two types of input: object-based and marker-based.

            \paragraph{Marker-Based Inputs} rely on tangible tokens that have a human-readable side for the players and a machine-readable side (i.e., markers) for the camera system (\autoref{fig:tangible-inputs}a, right).
            To give motion commands (i.e., to move player figures), players lay out movement tokens on the camera-based input plate (\autoref{fig:tangible-inputs}b).
            Depending on the character???s equipment, up to 4 such tokens can be placed and composed into a path command.
            The order of tokens (left$\rightarrow$right, top$\rightarrow$bottom) is taken into account to generate a path.
            Invalid commands (e.g., passing through walls) are discarded by the system.
            When confirmed by the player, \system detects their figure, grabs it, moves it along the selected path, and sets the figure back down at the target field.
            For interactions with NPCs, behavior tokens are used: they can be seen as ways to select specific options in interactions with the environment.
            
            \paragraph{Object-Based Inputs} Lastly, there are inputs that can not be represented with fiducial markers, either because the input space is too large (e.g., unique, crafted objects) or because they might compromise established visuals (e.g., dice), and are handled through contour-extraction (\autoref{fig:tangible-inputs}).
            For combat, dice are used, as is established in tabletop games (\autoref{fig:tangible-inputs}a, left). 
            Players may roll their dice--in the case of \system, six-sided--on the camera-based input plate, and the system detects the dice values using blob detection (while inverting the input to represent what the players see, i.e., the opposite side of the dice).
            Weapons and riddle solution attempts are similarly evaluated by the system (\autoref{fig:tangible-inputs}c).
            To do so, \system uses OpenCV (specifically the \texttt{cv.matchShapes()} function) to extract contours and compare them with a baseline representation (\autoref{fig:tangible-inputs}d).
            A weapon input is compared to all ``baseline contours'' (i.e., an axe, a sword, and a bow) to determine what the user has likely built.
            The lower the score, the higher the similarity.
            To determine the rarity, and, therefore, the power, of a weapon, the similarity is compared to quantiles determined in an informal study done during the design process (cf. \autoref{sec:designprocess}).
            The greater the similarity to this regular and symmetrical ``baseline contour'', the higher the rarity and the better the gameplay-relevant values of the weapon.

        \subsubsection{Projected Game Board}
\begin{figure}[h!]
          \centering
          \includegraphics[width=\minof{\columnwidth}{0.6\textwidth}]{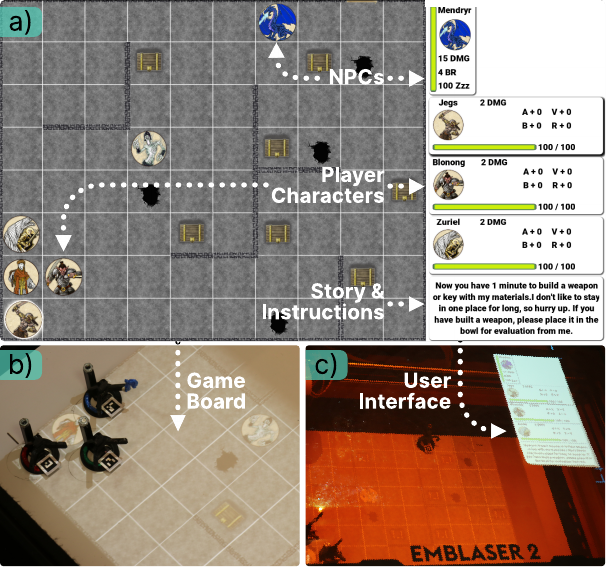}
          \caption{Overview over the projected output of \system: a)~output from the Godot application; b)~view inside the laser cutter with player figures on top of the projection (projection surface 1); c)~game UI with player and enemy statistics (projection surface 2).}\label{fig:game-details}
          \Description{}
        \end{figure}

            \system relies on projection (spatial augmented reality \cite{bimberSpatialAugmentedReality2005}) to add a dynamic game board to the surfaces involved in the game (\autoref{fig:game-details}).
            The projection maps to two depth levels, projection surfaces 1 and 2.
            Projection surface 1 is on the lid of the laser cutter (\autoref{fig:game-details}c).
            Here, the user interface is displayed, containing information about non-player characters, the players' characters, and narrative information (e.g., parts of the story, dialog elements, or instructions on what players have to do and whose turn it is).
            Projection surface 2 is inside the laser cutter (\autoref{fig:game-details}b).
            An MDF (medium-density fibreboard) plate with a thickness of 3mm is covered with a white sheet of glossy paper, which serves as the projection surface inside the laser cutter.
            Here, the game's board is displayed, and the figures are moved around through the gripping system attached to the laser cutter head.
            The Emblaser 2 laser cutter uses an orange-tinted laser safety glass in the lid.
            To counteract this optical alteration, a shader was written to accommodate for the induced tint and to retain contrast when players view the projection through the lid.
            This shader is only applied to projection surface 2, displaying the game board, and not to projection surface 1, where the UI (user interface) is displayed.
    
        \subsubsection{\system Game}
            The game part of \system was implemented using Godot\footnote{\url{https://godotengine.org/}, Accessed: 27.01.2024} 3.2.2, an open-source game engine. 
            Given that \system required 2D graphics (cf. \autoref{fig:game-details}), this proved to be a reasonable choice over alternative approaches.
            The laptop running the game uses an NVIDIA GTX 1060 graphics card and a 4-core Intel i7 processor. 
            We outline the specific gameplay elements in the following sections.
    
    \subsection{Gameplay}\label{sec:game}
        
        \system provides a collaborative tabletop scenario, meant to highlight the possibilities of hybrid board games that engage with and have control over physical/tangible elements of the play space.
        The following sections outline aspects of the game implemented and used in the study.
        
        \subsubsection{Introduction and Narration}
            \system supports narration in the form of spoken audio (in our case, spoken by a narrator), but also through textual output on the UI projected onto the laser cutter's lid.
            Users initially receive a longer introductory narration to present them with the game???s story and goals. 
            The narration then calls for the first actions to be performed by the players, turn by turn.
			Some of the instructions are timed, like fabricating something under a time limit.
            This applies to crafting weapons, and is explained through the merchant being impatient.
            Gameplay-wise, this creates a challenge and a sense of urgency.
            The narration component further provides dialog from NPCs (e.g., the merchant) and descriptions of events, like combat losses and wins, and damage taken or dealt.
            Dialog from different characters may also rely on a different voice, while background music and sound effects can be played for added immersion.
    
        \subsubsection{Scenario Goals}
            The goals of the implemented game follow the trope of ``escaping a dungeon'', where the players awake imprisoned, and have to collaborate to escape successfully.
            To do so, they need to forge\footnote{Both in the ``counterfeit'' and in the ``make''/``build'' definitions of the word.} a key. 
            Hints as to how the key looks are scattered across the map.
            Some can be won from the jailer, and others are found in chests.
            The implicit second goal is survival.
            Both goals require collaboration between the three players: combat is easier with multiple participants, and the hints can be found more quickly if the players spread out to search chests.
            
        \subsubsection{Game Board}
            The game board mimics the look of maps used for tabletop role-playing games like Dungeons and Dragons~\citegame{dnd}.
            These are often hand-drawn -- however, nowadays, there is a rich landscape of tools to support this.
            It consists of floor tiles, combining impassable terrain (holes, walls), interactive objects (the dungeon exit, chests), and characters (player characters and NPCs).
            Characters are digitally represented using circular sprites, with the currently active player receiving an outline around the respective sprite on the game board and the UI.
            \textbf{Movement across the board} is based on tokens (cf. \autoref{fig:tangible-inputs}b).
            Depending on the players' equipment (i.e., specific weapons increase or decrease their movement range), they can set up to 4 movement tokens per turn, which are converted to a movement command executed by the laser head in the order the players place them.
            When ending a movement next to an NPC, or when an NPC ends their movement next to a player, \textbf{combat or interactions} are triggered.
            Combat requires players to roll 2 dice, whose values are compared to the NPCs' digital dice roll.
            As with movement, weapons can grant players a round bonus, allowing them to roll the dice several times per combat.

        \subsubsection{Non-Player Characters}
            The game we implemented relies on a set of non-player characters (NPCs) to facilitate the envisioned gameplay elements (exploration, combat, crafting).
            The NPCs and their properties are outlined in the following paragraphs.\\
            
            \paragraph{Merchant} 
                The merchant is the first non-player character the players engage with.
                This NPC allows players to craft their first weapon, right at the start of the game.
                Seeking out the merchant during gameplay allows players to craft additional weapons for their figure.
                Doing so is likely to benefit their characters' properties and may make combat easier.\\
                
            \paragraph{Enemies}
                Combat is a mechanic commonly found in tabletop role-playing games.
                To enable this, \system provides a set of enemies for players to either actively engage with, or to avoid.
                A dragon is present in the upper part of the play area.
                It sleeps initially, but movement within its hearing range of 3 tiles will gradually wake it up.
                Players can circumvent this by ``sneaking'' past it (i.e., making one-tile movements per turn).
                The enemy dragon is an embodied NPC and has a tangible representation on the game board.
                When awake, it will be moved by the laser head toward the player figures and engage in combat.
                Additionally, weaker, non-tangible enemies are present: spiders, which spawn in regular intervals out of holes in the floor.
                While they are low-health, compared to the players or the dragon, they can still deal damage to unlucky (i.e., with their dice rolls) players.\\
                
            \paragraph{Jailer}
                In the narrative of the game, the players awake in a dungeon with no escape.
                The ghost of a jailer roams the dungeon and stands between the players and their escape.
                Unlike the enemies mentioned earlier, no combat is possible with the jailer.
                He is able to provide several hints about the key the players need to escape.
                When met, users can choose a behavior (aggressive, defensive, deceitful) to use when interacting with him.
                This represents a round of ``rock-paper-scissors'', with the jailer randomly picking an option.
                If the player wins, they get a hint or a different reward (e.g., healing).
                In case of a draw, nothing happens, and the jailer moves to a different position.
                In case the player loses, a negative effect is applied (e.g., damage or the dragon being deprived of sleep points).
                After each such interaction, the jailer moves on and has to be reached by one of the players to trigger this minigame another time.

        \subsubsection{Craft Activities}\label{sec:crafting}
            \begin{figure}[h!]
              \centering
              \includegraphics[width=\minof{\columnwidth}{0.6\textwidth}]{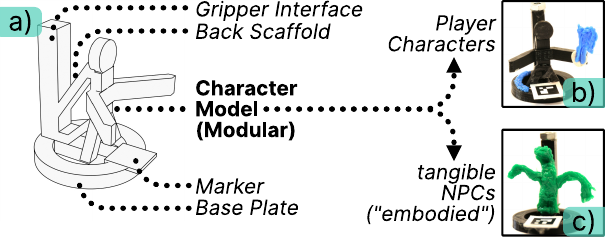}
              \caption{Modular character models consist of a base~(a) to enable tracking/gripping and enable different character designs, used for player models~(b) or tangible NPC figures~(c).}\label{fig:physical-parts}
              \Description{}
            \end{figure}
            
            Over the course of a \system game, players have to engage in a set of craft/fabrication activities with the 3D-printing pen.
            Initially, they may customize their game figure using the pen.
            \system relies on a standardized base design for figures 
            (\autoref{fig:physical-parts}a), which either provides a skeleton to customize 
            (\autoref{fig:physical-parts}b), or could leave room for design ``from scratch'' 
            (\autoref{fig:physical-parts}c).
            Each time players meet the merchant, they may fabricate one of three weapon types: a bow, an axe, or a sword.
			Players are provided with a rough guidance in terms of size and some examples of how their artifact may look like.
            The crafted weapons are put onto the input plate and are scanned and rated.
            The rating consists of video-game levels of rarity, ranging from common, over rare, to legendary.
            The type of weapon and its rarity influences the bonuses players receive from them (e.g., attack damage, movement range, or turns usable for combat).
            After that, layers retrieve their figure from the laser cutter and attach the weapon to it.\\

            When arriving at the dungeon door towards the end of the game, players can attempt forging the key they need to win their scenario.
            The solution consists of 3 elements: a trident and a circle, connected with a line.
            Each of these elements is conveyed through riddles (e.g., \textit{``One part of the key is the shape that does not fit into this row: G D B L C S''}\footnote{Solution: Here, the solution is ``L'', as that is the only letter in the row without curves. This is seen in \autoref{fig:studyartifacts}.1 b), where the ``L'' is the connecting element}).
            The key is evaluated in a similar fashion to weapons, with the output being success or not.
            An unsuccessful attempt draws the attention of enemies.
            This happens towards the end of the game and may be preceded by a longer discussion phase between the players, trying to combine the hints received over the course of the game into a (physical/spatial) solution.

     \section{Evaluation}
\label{sec:evaluation}
    To further explore how \system augments and alters the board game experience, we set out to test it with a set of participants.

    \paragraph{Study Design} 
        The goal was to evaluate \system in exploratory sessions focusing on players' holistic experiences.
        Therefore, the study was framed as a collaborative game session that happens to be preceded and followed by questionnaires.
        The study was similarly meant to confirm the functionality of \system when used by players and to explore technical limits.
        
    \paragraph{Metrics}
        The core (quantitative) metrics, on which we focused, were affective state (i.e., valence, arousal, and dominance measured using the Self-Assessment Manikin~\cite{bradleyMeasuringEmotionSelfassessment1994}), connection to one's figure (measured using the Inclusion of Other in the Self (IOS) scale~\cite{aronCloseRelationshipsIncluding1991, aronInclusionOtherSelf1992}, adapted to refer to the players' closeness with their figure), and a set of Likert-scale questions regarding sentiments towards fabrication technologies and the game mechanics that \system introduces.
        These metrics were measured before and after the game session to allow for comparisons.
        Lastly, we solicited open comments from the participants and qualitatively analyzed video recordings from the play sessions along with the observation notes that the instructor took during the study.
        
    \paragraph{Participants}
        We recruited 4 groups of 3 (triads) of participants around our institution.
        To avoid people feeling uncomfortable in the context of co-located, collaborative gaming, we ensured that the participants in each triad were at least familiar with each other and could imagine playing a board game together.
        This was addressed during recruiting.
        4 participants identified as women, 7 identified as men, and one participant identified as non-binary.
        Their age ranged from 24 to 29 ($M = 25.8$).
        All participants were either students or researchers and had played board games before. 
        In the last 6 months before the study, 6 reported having played over 12 times, and 6 reported having played between 1 and 6 times.
        6 participants reported having played purely digital board games before, and 3 participants had played hybrid board games.
        8 participants reported having used a 3D-printer before, 4 participants had used a laser cutter before, and only 3 participants had used a 3D-printing pen before.
        4 participants had no contact with any of these devices before.
        2 additional participants had only used 3D-printers 1-2 times before.
        We classify these participants with 2 or fewer exposures to 3D-printing as \textit{``novices''} to digital or personal fabrication \cite{hudsonUnderstandingNewcomers3D2016,bermanAnyoneCanPrint2020,stemasovRoadUbiquitousPersonal2021}, and people with more than 2 exposures to 3D-printing as \textit{``non-novices''}.
    
    \paragraph{Procedure}
        Participants initially received an introduction to the study process, provided informed consent, and filled out a first questionnaire.
        We started a video recording using a camera pointed at the play area as soon as the game began to allow for later analysis of events (\autoref{subsec:observations}).
        The three participants played one uninterrupted game session under the supervision of a study conductor.
        A game session lasted between 30-50 minutes.
        While there was no time pressure, the conductor increasingly provided hints (i.e., regarding strategy) on how to progress as time went on, to give all groups a chance to attempt to solve the riddle and succeed in the scenario.

        \subsection{Questionnaire Data}
        
        \begin{figure*}[h!t]
          \centering
          \includegraphics[width=\linewidth]{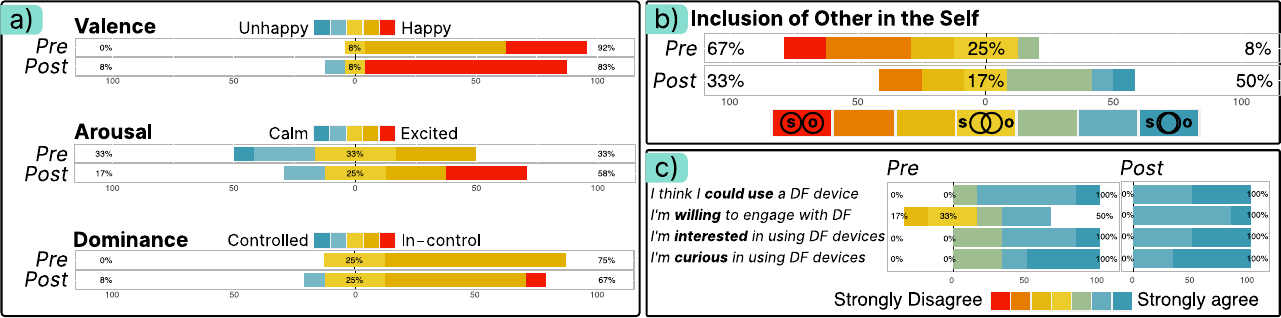}
          \caption{Answer distributions for a) the Self-Assessment Manikin (SAM); b) Inclusion of Other in the Self (IOS) Scale; and c) select, custom Likert-Scaled questions. All plots present answers before (''pre'') and after (''post'') the participants' game session.}\label{fig:plots}
          \Description{}
        \end{figure*}

\paragraph{Standardized Questionnaires} 
        For the Self-Assessment Manikin (SAM), participants were queried about their affective state before and after the game session (\autoref{fig:plots}a).
An exact Wilcoxon-Pratt signed-rank test showed that a game session of \system yielded no significant effect on participants' reported valence ($Z = 1.22, p = 0.28, r = 0.25$) between pre- ($ M = 4 $) and post-game ($ M = 5 $) and dominance ($Z = -0.13, p = 1, r = -0.12$) between pre- ($ M = 4 $) and post-game ($ M = 4 $) using the SAM.
However, the test yielded a significant difference between pre- ($ M = 3 $) and post-exposure ($ M = 4 $) responses for the arousal subscale of the SAM ($Z = 2.19, p = 0.047, r = 0.45$). 
For the Inclusion of Other in the Self (IOS) Scale, participants were also asked to ``... select the picture that best describes your relationship with your game figure'' (\autoref{fig:plots}b).
        An exact Wilcoxon-Pratt signed-rank test showed that a game session of \system yielded significant differences in the IOS-data between pre- ($ M = 2.5 $) and post-exposure ($ M = 4.5 $) responses ($Z = 2.41, p = 0.014, r = 0.49$).

        \paragraph{Ratings}
We further acquired general sentiments towards hardware and the notion of personal fabrication before and after ``exposure'' through the game.
            The statements were rated on 7-point Likert-Scales (a subset is seen in \autoref{fig:plots}c).
When analyzing all 12 participants, independent of experience level with fabrication devices, no significant differences were shown using the Exact Wilcoxon-Pratt Signed-Rank Test, except for the statement \textit{``I can imagine DF devices being used for tasks other than manufacturing things''} ($Z = 2.89, p = 0.0039, r = 0.59$) between pre ($ M = 5 $) and post-game ($ M = 6 $).
When focusing on the 6 participants that could be defined as novices (for 4 of them, \system was the \emph{first and only exposure} to personal fabrication devices), additional trends could be observed.
            An Asymptotic Wilcoxon-Pratt Signed-Rank Test indicated that there is a significant difference between pre- and post-game sentiments towards fabrication devices for the statements 
            \textit{``I am willing to engage with DF devices''} ($Z = 2.89, p = 0.083, r = 0.5, M_{pre} = 6, M_{post} = 6.5$) \textit{``I am interested in using DF devices''} ($Z = 2, p = 0.045, r = 0.58, M_{pre} = 6, M_{post} = 6.5$), and \textit{``I am curious in using DF devices''} ($Z = 1.73, p = 0.08, r = 0.5, M_{pre} = 6.5, M_{post} = 7$).

        \paragraph{Mementos}\label{par:mementos}
When asked which tangible artifacts they would want to keep after their (or comparable) game session, 8 participants reported wanting the figure (and their weapon), 7 would have liked to keep the key, 5 were interested in NPC figures, 3 were specifically interested in damaged artifacts (e.g., figures), and 2 would have liked to keep the game board, which gathered new scuffs each play session.
    
        \paragraph{Experience}
Before experiencing \system, participants were asked a set of Likert-scale questions about prospects regarding the scenario.
            67\% expressed reservations regarding their own figure being damaged, yet only 25\% felt the same way regarding their teammates' figures.
            The laser cutter being responsible for motion (instead of the players themselves) was appealing to 75\% of the participants.
After the play session, participants rated a set of Likert-Scaled questions regarding aspects of \system again.
            75\% agreed that the potential of damage improved their experience, 92\% wanted to fight more to see damage done to the NPC figure, 83\% rated the combination of board game and laser cutter as an ``improvement to the game experience'', and all participants considered the crafting activities with the 3D-printing pen as a similar improvement.
            However, several comments were raised regarding that they would have welcomed more frequent crafting activities, which we deliberately limited to ensure a compact study session.

    \subsection{Observations and Open Comments}\label{subsec:observations}
        \begin{figure}[h!t]
          \centering
          \includegraphics[width=\minof{\columnwidth}{0.6\textwidth}]{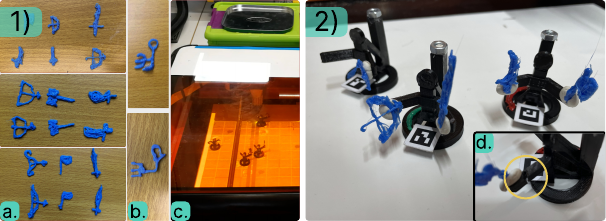}
          \caption{Subset of artifacts made by the participants during their playthroughs. 1)~Crafted objects (a.:~weapons; b.:~riddle solutions, top--incorrect, bottom--correct; c.:~late-game state of a group, seeing a cut process).~2) Figures as customized by participant group 3, with cutting damage visible~(d.)}\label{fig:studyartifacts}
          \Description{}
        \end{figure}

        We further analyzed the observation notes, collated the open comments, and identified common motifs occurring in the video recordings.
        A set of crafted artifacts from the study can be seen in \autoref{fig:studyartifacts}.
        3 out of 4 groups experienced the laser cutter damaging/cutting one of the figures (cf. \autoref{fig:studyartifacts}c and d). 
        Similarly, 3 out of 4 groups succeeded in solving the riddle and thereby won the scenario.

        \paragraph{Video Analysis Process}
            All game sessions were recorded for later analysis.
            We followed a reflexive thematic analysis approach to understand behaviors and patterns better.
            The approach can be described as \emph{inductive}, as we particularly focused on player experience and how it changes through the hardware and software setup of \system.
            The audio files were first machine-transcribed on a local system using whisper.cpp\footnote{\url{https://github.com/ggerganov/whisper.cpp}, Accessed: 28.01.2024}. 
            One of the authors then reviewed all recordings, correcting the transcript where necessary. 
            This was, for instance, the case when several people were talking simultaneously. 
            At this point, annotations for \emph{machine actions} (e.g., motion, cutting, or narration elements), and crafting actions (e.g., crafting the key) were added to the transcript.
            Two of the authors then, independently of each other, watched all recordings and created separate codebooks (21 entries and 47 entries, respectively).
            The two codebooks were then merged into a single one, leaving 46 entries grouped into 8 high-level topics.
            The merging process required the 2 authors to present examples and collaboratively merge and group the entries.
            The merging process was also used to sharpen individual code formulations (e.g., \textit{``Skill judgment of using fabrication tools based on prior usage''} $\mapsto$ \textit{``Craft-Self-Assessment''}). 
            Sample quotes of the identified categories were then extracted and assigned labels accordingly.

        \paragraph{Behaviors}
            The overall topics from the video analysis are briefly outlined in the next paragraphs.
            \textbf{General Player Experience(s)} consists of acts like encouraging peers or visible enjoyment (e.g., laughter, celebration, and friendly banter among teammates).
            This happened, both during crafting, but also during gameplay itself, with each group voicing desires to keep playing beyond the planned game duration. 
            Solving the riddle at the end and solving individual clues also sparked enthusiasm (e.g., \iquote{Print! Print! Print!}{P4} -- when P6 was fabricating the key).\\

            The overall player experience is connected to notions of \textbf{Collaboration} present in \system.
            All participants engaged in pointing and deixis for collaboration and planning.
            Most pointed at figures and the board themselves, while P8 leveraged the projection to precisely point using their finger's shadow.
            Players kept strategizing together and acting as a team to move towards their common goal, and engaged in task division on the game board (e.g., players with more movement range being delegated to opening all chests, and players with more attack power doing combat) but also outside of it (e.g., people waiting for their turn trying to solve the riddle, while others gathered additional clues).\\
            
            \textbf{Gameplay} aspects included players trying out individual features (e.g., motion, tokens) or complaints about rewards or randomness (P8, P9, P11)
            Participants also referred to other games to reason about how \system may behave (e.g., whether running away allows NPCs to still hit one's figure -- P1).
            This is linked to participants trying to manage their expectations towards \system. 
            For instance, this includes participants expecting the spider to drop loot to improve their equipment: \iquote{drops [...] poison or something that you could use on an enemy}{P12}, \iquote{Poison arrows! [referring to P12 carrying a bow]}{P10}.\\

            The aforementioned facets of gameplay connect to aspects of \textbf{Narrative, Story, and Immersion}, where participants engaged in roleplaying and voiced expectations of the game world and logic.
\system already provides some degree of narration to set the stage, but participants tried to weave their own stories during gameplay, as is usual in tabletop RPGs~\cite{whitePlayerCharacterWhatYou2014}: 
            \iquote{I'd like to escape ... but everyone's in my way}{P2},
            \iquote{Excuse me? I'm fighting heroically here}{P3}.
Then again, there were also breakdowns: \iquote{He [the jailer] is totally in my way [...] he's a ghost, why can't I move through him?}{P1}, where the absence of a physical NPC (cf. \autoref{fig:physical-parts}) and their suboptimal digital representation (i.e., a ghost-like sprite on the board) lead to incorrect expectations.\\
            
            Regarding \textbf{Hardware}, various enthusiastic reactions to robotic motion were present across all participants. 
Statements like \iquote{He'll get it [the figure]. Yeah, whoo}{P2} highlight not only a degree of acceptance of re-assigned roles (cf. \autoref{sec:formativesurvey}), but also users enjoying the robotic motion (\iquote{(cutter moves) [...] magic! (laughter)}{P10}.
            This also applied to technical details, like the gripping system's correcting motions after recognizing the marker (\iquote{(laser corrects and grips) [...] Wonderful!}{P4}.
This also includes humming an ominous melody\footnote{In the case of one participant, this was the Imperial March from Star Wars.} when the laser cutter made its move, talking to the fabrication devices (P8, P12), or imitating their sounds (P3, P7).
            As with aspects of world-building mentioned earlier, participants repeatedly voiced questions regarding rules, gameplay, or hardware.
For example, P8 wanted to ensure that their 4 movement tokens ($\uparrow,\rightarrow,\downarrow,\downarrow$) translate to the right path around a wall: \iquote{Does it ''get'' their order? (points to input and movement tokens)}{P8}.\\
            
            \textbf{Crafting and Equipment} are crucial elements of \system, given that outcomes of the crafting activity \emph{influence gameplay} and are not only cosmetic.
            This led players to self-assess their skills or to assess others' skills through the lens of \system's ratings or alter their strategy accordingly (e.g., people with a movement bonus focusing on exploring the map).\\
            
            \textbf{Damage and Combat} was another core aspect of \system. 
            Here, participants embraced this notion and either \emph{avoided} combat (\iquote{[...] (Points on Map) should I really go past the dragon?}{P11}), \emph{provoked} it (\iquote{Don't go to the exit just yet, I want to beat him [the dragon] up.}{P12 to P10}), or considered it an intriguing experience (\iquote{(to the instructor) can you do it again?}{P9, after seeing damage happen}).
            This is linked to groups coordinating their efforts based on equipment (which influenced stats like attack power, but also movement range).

        \paragraph{Open Comments} 
            Participants were also asked to leave comments in the questionnaire regarding their experience, which were collated by the authors.\\
            
When asked about \textbf{positive aspects} of their experience with \system, participants highlighted their enjoyment~(6 mentions out of 12 -- 6x), the crafting activities~(8x), the aspects of collaboration~(3x), novel technical aspects (e.g., machine-driven motion, 8x) and \system's narration~(2x).
            Participants also appreciated the prospect of ``real consequences'', both with respect to destruction/damage, but also with respect to the machine moving tangible elements of a game or the crafted weapons having an impact on the game (e.g., depending on their rating).\\
            
Participants were asked to describe \textbf{potential improvements} about the system and their experience.
            A recurring topic was the pace of the game, as giving motion commands was perceived as slow, and so was the execution of the motion commands by the laser cutter.
            Gameplay-wise, players mentioned potential improvements regarding the balancing choices (e.g., chest contents) or lower randomness during the ``combats'' with the jailer.
            Players also mentioned other considerations, such as the volume of the laser cutter when thinking about home use~(1x), occlusion of the projection~(1x), and the aspect that fabrication/craft activities were too rare~(1x).\\
            
When asked to describe their \textbf{experience in general}, participants repeated previous notions of enjoyment and fun~(9x), voiced wishes to re-play the same scenario (1x), highlighted the positive effects of the narration~(1x), and mentioned a potential for long-term motivation~(1x).
            4 participants further mentioned the social, collaborative nature of the game as a positive and engaging factor.

 \section{Discussion}
\label{sec:discussion}
    The following sections first discuss the study results and then consolidate them---together with insights from the design and development process of \system---into a set of design considerations for fabrication-augmented hybrid board games.

    \subsection{User Study}
        Through the user study, we uncovered several aspects that not only confirm the functionality of \system as a hybrid game but also aspects that indicate feasibility and value beyond the specific game that was implemented.

        \paragraph{Avatar Customization: }
            A session of \system increased the players' connection with their figures, as measured through the IOS questionnaire.
            Notably, players could only personalize their characters through crafted equipment, and not by altering the figures themselves due to time constraints in the study.
Despite this, participants enjoyed self-expression by crafting the weapons for their figures themselves.
            In comments, they described this self-expression, which had functional effects on the game, as fun and exciting.
            Players also grew attached to these artifacts, not wanting to lose them during the game (\iquote{Noo, don't kill my weapon}{P9}\footnote{in response to the laser head accidentally catching on their figure, toppling it}), but also expressing interest in keeping them beyond the session as mementos (cf. \autoref{par:mementos}).
This notion can undoubtedly be pushed further (e.g., customizing the avatar itself, potentially with tools other than the 3D-printing pen), and explored through lenses of user representation~\cite{seinfeldUserRepresentationsHumanComputer2021} and identification~\cite{burukWEARPGMovementBasedTabletop2017, kaoEffectsBadgesAvatar2018}.
            Yet, this already highlights the potential of fabrication to strengthen the players' connection with their figures by replicating customization concepts known from digital games~\cite{cuthbertEffectsCustomisationPlayer2020,turkayEffectsCustomizationMotivation2015} in the physical domain. 
            
        \paragraph{Player Experience:}
            Overall, \system provided a positive experience for the players. 
            Considering valence and dominance, the responses increased, but without a significant effect. 
            Given the high responses before play and our low sample size, we cannot draw clear-cut conclusions but see trends of a positive player experience that require further work to confirm effects.
            However, participants mentioned various reasons for positive valence (e.g., winning, collaboration, novel experiences) and despite \system taking a core mechanic (i.e., moving) away from the users, and deciding their order of actions, participants did not report a lower dominance after the play.
            Regarding arousal, players showed an increase after the experience, highlighting the engaging nature of play. 
            The differences in arousal may also indicate a positive game experience (if not aimed at relaxation), i.e., an exciting experience that could be associated with a state of flow \cite{nackeFlowImmersionFirstperson2008}.
            This is also evident in the open comments, where participants described the game as \textit{``fun and immersive''}, \textit{``a fun group experience''}, or that it was \textit{``fun to design weapons and interact with the environment through the system''}, for example.
            Players specifically positively mentioned experiences of uncertainty (e.g., regarding weapon ratings and NPC behavior), curiosity (e.g., regarding the technology and scenario progression), and autonomy (e.g., strategizing together), concepts that have been shown as essential for engaging play \cite{costikyanUncertaintyGames2013, powerLostEdgeUncertainty2019, toIntegratingCuriosityUncertainty2016}.

        \paragraph{Jealousy and Benevolence: }
            Craft and subsequent machine-driven assessment of it lead to reactions from teammates.
            We generally observed positive reactions between players, who seemed to emphasize the collaboration aspect of the game. 
            For example, in response to P10 crafting a weapon: \iquote{oooh, a rare bow!}{P11},
            \iquote{Now you're competing with me!}{P12},
            \iquote{Now the party is all geared up.}{P11}.
Praising teammates' craft skills, as rated through \system, was also done regularly, and participants with imperfect outcomes even attempted to defer activities to more ``skilled'' teammates: \iquote{Can I have [P7] build a weapon for me?}{P8}.
            These aspects highlight how craft, followed by machine-driven assessment, may affect and influence players, their perceived proficiency, and game strategies.
            Granted, this notion is particularly dependent on the players' personal relationships and individual characters, which manifest as player types.
            
        \paragraph{Idleness and Waiting: }          
            In its current implementation, \system is arguably a slower experience, as actions like motion are mediated through the machine for safety reasons. 
            Despite this slow turn-taking, no signs of impatience were seen.
            Damaging was particularly slow, yet players' curiosity outweighed this downside.
            This can be linked to the work of Tanenbaum~\cite{tanenbaumDesigningHybridGames2017}, who outline \emph{idleness} as a design element. 
            Using faster (i.e., laser-cutters~\cite{hellerCutCADOpensourceTool2018}) or less precise (i.e., 3D printing pens~\cite{takahashi3DPen3D2019}) accommodates for slowness, but does not remove it entirely.
            For \system, we noticed that there was either excited observation of machine action, or, alternatively, participants strategizing in parallel.
 
        \paragraph{Stances towards Fabrication Hardware: } 
            Apart from acknowledging that fabrication hardware could be employed for tasks other than explicit fabrication, there were no significant differences regarding our single-item questions.
            Arguably, our sample contained a higher-than-average amount of users experienced with fabrication devices.
            When zooming in, trends among novices became visible: novices did--albeit slightly--report a more positive stance (i.e., interest, curiosity) toward such devices.
            While \system does not aspire to be an explicit teaching and learning tool or environment, it still managed to spark joy across all players and curiosity across the players new to personal/digital fabrication. 
            Notably, it did so without requiring users to ``learn'' or ``design'' or ``iterate'', but instead letting them express themselves through the lens of a hybrid board game that engages not only additive and subtractive manufacturing but also robotic motion (i.e., CNC, as used in pick-and-place machines).
            We believe that this indicates a new way to make ``first contact'' with the technology, preceding explicit learning activities while being an enjoyable way to engage with devices.
            This approach is promising, but needs more thorough exploration through studies.

    \subsection{Design Considerations}
        The following sections outline a set of design considerations that we have derived based on the design and development process of \system, our formative survey, playtesting, and the user study we conducted.
        See \autoref{tab:designconsiderations} for an overview.

        \begin{table}[h]
            \caption{Overview of the design considerations based on the design and development of \system and the conducted studies.}
            \label{tab:designconsiderations}
            \centering
\resizebox{\minof{\columnwidth}{0.6\textwidth}}{!}{\begin{tabular}{p{0.07\linewidth} p{0.93\linewidth}}
                \toprule
                & Design Consideration\\
                \midrule
                \hyperref[par:agency]{\textbf{DC1}} & \textit{Foster Player Agency through Physical Expression}\\
                \hyperref[par:riskreward]{\textbf{DC2}} & \textit{Balance Risk and Reward}\\
                \hyperref[par:continuity]{\textbf{DC3}} & \textit{Embrace Continuity in Hybrid Board Games}\\
                \hyperref[par:sustainability]{\textbf{DC4}} & \textit{Consider Sustainability when Damage Becomes a Design Element}\\
                \hyperref[par:transparency]{\textbf{DC5}} & \textit{Leverage System Transparency vs. Opaqueness}\\
                \hyperref[par:playfulfab]{\textbf{DC6}} & \textit{Facilitate Playful Engagement with Fabrication Devices}\\
                \bottomrule
            \end{tabular}
            }
        \end{table}
           
\paragraph{Foster Player Agency through Physical Expression: }\label{par:agency}
            Agency perceived by the players is a crucial aspect of any game experience.
            \system added a new aspect of control and self-expression: namely, the crafting of tangible artifacts that have a meaningful (i.e., gameplay-relevant) influence over the game progression.
            For \system, we had to balance these aspects with users' desires to engage in a board game experience they might be more familiar with: one where they have full control over logic, narration, and physical elements.
            The slowness and occasional unreliability of motion further drew attention to the notion of agency and control.
            \system added and provided novel elements for a hybrid board game experience, but also took away elements that are an integral part of board games (e.g., moving figures, in our case, for safety and technical reasons).
            We argue that such tradeoffs in re-shuffling chores~\cite{xuChoresAreFun2011} or even fundamental interactions like figure movement can be worthwhile, if \emph{agency is fostered through other means}, such as physical creation mechanics.

\paragraph{Balance Risk and Reward: }\label{par:riskreward}
            The notion of risk and risk-taking behavior is already present in most games.
            However, their consequences differ in severity: they may apply to a single turn, a game session, or, in more extreme cases, the loss of characters in which players have invested months or years of their lives~\cite{abrahamImposedRulesExpansive2013,westItAllFun2022}. 
            This behavior is likely altered further when consequences are physical and tangible.
            In our study, we observed both combat-avoidant behavior and combat-provoking behaviors, depending on the equipment players possessed,their stance towards loss (cf., loss aversion in games~\cite{bandeiraromaotomeRiskingTreasureTesting2020}), or their curiosity to \emph{see} a previously unseen and unkown~\cite{toIntegratingCuriosityUncertainty2016} event.
            In the case of \system, the rewards, beyond the experience of combat itself, were unclear.
            This was indicated by participant comments, calling for more \emph{tangible rewards} of combats while the rewards in \system were immaterial (i.e., hints to solve riddles or \emph{digital} boons like healing). 
            We argue that material risk may need a material reward, which happens to be well within the realm of fabrication-augmented hybrid board games.
            It is unlikely that figures with financial values found in ``Warhammer'' would even be used in scenarios where destruction is an option.
            Similarly, the \emph{absence} of material risks on persistently used figures may not yield altered gameplay strategies, as observed in our study.

        \paragraph{Embrace Continuity in Hybrid Board Games: }\label{par:continuity}
Games like Dungeons and Dragons and Legacy games~\cite{reiberMajorDevelopmentsEvolution2021} (introduced with ``Risk Legacy''~\citegame{risklegacy}) embrace continuity that carries over several game-sessions, where players develop backstories and increasingly improve their characters~\cite{arnaudoStorytellingModernBoard2018}; where new rules are introduced, or game boards change dynamically.
With approaches like \system, owning a figure for board games may become even more relevant: players may re-use the same figure (i.e., their avatar), or re-use a game board, physically marked with traces of previous adventures or encounters, building a collaborative, tangible memento~\cite{turakhiaFabOIntegratingFabrication2021,jonesWearYourHeart2023}, which may have a rich, gameplay-affecting (digital) component.
            In turn, the gradient between physical alteration, damage, and destruction has to be considered carefully during game design to balance risk and meaningful physical continuity, as mentioned in the previous point.

        \paragraph{Consider Sustainability when Damage Becomes a Design Element: }\label{par:sustainability}
In continuation of the two previous points, this gradient between physical ``alteration'', ``damage'', and ``destruction'' also has to be carefully considered with respect to material use and sustainability.
            This consideration can be woven into aspects like material choice \cite{riveraDesigningSustainableMaterial2023,vasquezMycoaccessoriesSustainableWearables2019,stemasovEphemeralFabricationExploring2022}, but we argue--grounded in our own design process--that it can not be omitted from game design itself, considering it upfront and not as an ``afterthought''~\cite{stemasovEphemeralFabricationExploring2022}.
            A middle ground has to be found, where actual ``destruction'' is rare, while retaining risk yet demonstrating it through visible, tangible damage.
            This echoes notions and design considerations around ``permadeath'' in digital games \cite{carterFearLossMeaningful2017}.
            The consideration of sustainability can also manifest in re-usable figure designs, as done with \system, to minimize damage, or embed \emph{repair}~\cite{jonesPatchingTextilesInsights2021} in the fabrication-gameplay.
            For instance, we see the damaged (not broken) figure in \autoref{fig:studyartifacts}.4 to be one that players could mend with a 3D-printing pen and keep using it as a ``battle-tested'' character, specifically \emph{because} of the visible, repaired damage.
            In the study, such damage was labeled by the participant to be a \iquote{[...] Hero's scar}{P12}.

\paragraph{Leverage System Transparency vs. Opaqueness: }\label{par:transparency}
            In the case of \system, players were not always content with how their weapons were rated, given that little explanation was provided. However, at the same time, this unpredictability of \system was appreciated and valued, as it seemed to be ``truly impartial'', akin to a genuinely experienced and competent game- or dungeon-master in pen-and-paper roleplaying games.
            This echoes notions of uncertainty in game experiences~\cite{costikyanUncertaintyGames2013, toIntegratingCuriosityUncertainty2016} and can be further augmented with uncertainty in creation (e.g., quality assessment) and loss (e.g., degree of damage).
            While opaqueness is generally not ideal in game design, a Dungeon Master's plans for a campaign are part of a positive experience~\cite{acharyaInterviewsDesigningSupport2021} -- \system tries to embody this by controlling NPCs, or evaluating crafted artifacts.
            In contrast, participants complaining about randomness (e.g., \iquote{It doesn't matter at all which number you get, it's all completely random [...]}{P8}\footnote{This referred to the combat, which for both NPCs and players is based on random dice throws}) could be traced back to an excess of opaqueness in system-controlled moves. 
            Showing random values of digital dice throws (e.g., as done by the NPCs) would have prevented frustration.

\paragraph{Facilitate Playful Engagement with Fabrication Devices: }\label{par:playfulfab}
            Personal fabrication machinery is generally complex and has sparked a variety of attempts to make it more accessible (e.g., \cite{baudischKyub3DEditor2019,muellerConstructableInteractiveConstruction2013}) and widely adopted (e.g., \cite{stemasovRoadUbiquitousPersonal2021,follmerCopyCADRemixingPhysical2010,tianMatchSticksWoodworkingImprovisational2018}).
            Usually, ``\textit{first contact}'' happens through established design processes (i.e., solving a mechanical problem or designing something appealing~\cite{liGamiCADGamifiedTutorial2012}).
            We argue that this approach may be changed by \emph{remixing the way we use and engage with these devices}: by omitting the notion of design and providing a playful, immersive engagement with the technology instead of focusing on problem-solving upfront~\cite{stemasovEnablingUbiquitousPersonal2021}.
            The potential for this lens is indicated by the responses given by novices in our study, but also in related works~\cite{lafreniereBlockstoCADCrossApplicationBridge2018,follmerKidCADDigitallyRemixing2012}, especially ones that explore the use of fabrication technology \emph{for} board games \cite{bhaduriFabricationGamesUsing2017,tanenbaumDesigningHybridGames2017}.
            
    \subsection{Emergent Design Elements Beyond \system}\label{sec:emergentelements}
        \system leverages a unique hardware setup to provide an engaging experience while addressing the design goals presented in section \ref{sec:designgoals}.
        However, the fundamental concepts likewise apply to alternative game genres and alternative hardware setups.
        By combining the individual components of \system (cf. \autoref{fig:teaser}), namely creation, input+assessment, motion, destruction, output+narration, a \emph{rich space of altered, augmented, or re-assigned roles} and mechanics in hybrid board games emerges.
        We considered these combinations of emergent design elements to be a design exercise, akin to a morphological analysis~\cite{zwickyMorphologicalApproachDiscovery1967} to describe and outline the potential of comparable setups through the lens of existing board games.
        These combinations may exhibit effects in terms of technological possibilities (\emph{enabling} interactions), but also \emph{alter} underlying relationships between users, game artifacts, the game, and devices involved~\cite{devendorfBeingMachineReconfiguring2015}.
        The following examples, in no particular order, touch upon existing traditional and hybrid board games but could potentially be used to generate novel game designs and platform configurations.
        We consider these brief examples to highlight the richness of the space we probed with \system, and consider the exploration of this space to be a highly promising direction for future work. 
        
        \paragraph{Creation + Sensing} 
            Combining creation and sensing enables assessing player-crafted artifacts and assigning game-relevant stats, potentially facilitating agency, attachment, and value-creation. 
            This considers the opposite direction of FabO~\cite{turakhiaFabOIntegratingFabrication2021} by weaving analog craft into digital parts of gameplay while also \emph{assessing} it for the game environment and its narrative.
            A tighter coupling between physical and digital elements in hybrid board games is enabled by such connections (physical $\longrightarrow$ digital).
            For instance, an intricate hotel in ``Monopoly''~\citegame{monopoly} may yield extra income, or a particularly well-painted miniature figure used in ``When Darkness Comes''~\citegame{whendarknesscomes} receiving a bonus to their charisma value. 
            
        \paragraph{Motion + Sensing} 
            Motion and sensing enable the detection and movement of figures or other elements on the board.
            This enables digital game logic to have effects on tangible components (digital $\longrightarrow$ physical), but also allows the game to enact players' motion commands (as done in \system), or acquire and react to player-initiated state changes on the board (physical $\longrightarrow$ digital). 
            In turn, this may enable tangible remote play~\cite{maurerDislocatedBoardgamesDesign2019,millsDistributedLetterJam2021, millsRemoteWavelengthDesign2023,sparrowLessonsHomebrewedHybridity2023}, as demonstrated by commercial projects like Phantom Chess\footnote{\url{https://www.phantomchessboard.com/}, Accessed 25.01.2024}.
            
        \paragraph{Creation + Destruction} 
            Allowing creation in conjunction with the risk of destruction adds the potential for permanent alteration and loss of self-crafted objects, which, in turn, adds risk and altered risk-taking behavior.
            This further goes beyond the destruction of ``owned'' objects \cite{eickhoffDestructiveGamesCreating2016}, towards self-crafted, and likely valued, ones~\cite{rossmyPointNoUndo2023}.
            For miniature wargames like ``Warhammer''~\citegame{warhammerthirdedition}, this may mean that players' self-painted figurines weather over time (e.g., from combat or environmental effects), or that particularly risky decisions may have lasting consequences to artifacts players have grown attached to.
            This bridges decisions that may emerge from in-game calculations to physical outcomes (digital $\longrightarrow$ physical).
            
        \paragraph{Sensing + Destruction} 
            Combining sensing and destruction allows digital components of the game to detect player-made damage or react to inputs with damage to physical components (e.g., after a loss in combat).
            This may also allow iterative sensing after destruction to meaningfully assess the damage and understand the random effects of these actions (e.g., whether the laser was focused enough to damage a figure or not). 
            In the study, there was an instance of the cut stopping too early, which made for an instance of \emph{damage} over \emph{destruction}, which could have nuanced effects on gameplay (physical $\longrightarrow$ digital $\longrightarrow$ physical).
            For single-use games like Exit~\citegame{exitthegamethesecretlab}, a system may sense whether players are about to enact the ``correct'' destruction of physical game materials, or embed the sensing and verification of destroyed elements into the gameplay. 
            
        \paragraph{Sensing + Output} 
            Diegetic feedback for crafted artifacts is enabled by combining sensing and output.
            Then, real craft translates to in-game forging mechanics (physical $\longrightarrow$ digital), as presented with \system. 
            In games like ``Catan''~\citegame{thesettlersofcatan}, player-crafted tiles could be automatically assigned a resource type (forest, hill).
            Furthermore, this covers the automation of \emph{any} translation of physical elements to digital ones.
            Users are then freed from--often manually--keeping a digital representation of a game in sync with the physical one.
            
        \paragraph{Motion + Output} 
            Motion coupled with output allows for narrations and explanations of NPC actions, further supporting an immersive experience in narrative-driven board games (e.g., for alien non-player characters in ``XCOM the Board Game''~\citegame{xcomtheboardgame}).
            This can serve as a narrative bridge (physical $\longleftrightarrow$ digital) that provides explanations of actions in the physical space, both for NPC actions, but also to narrate and enrich players' actions, serving as a game master, as demonstrated with \system.\\

            This subset of design element pairings already outlines a highly intriguing potential for altered and enhanced player experiences.
            However, they are certainly incomplete and may develop different dynamics depending on the specific game and hardware platform they are applied to.

             \section{Limitations and Future Work}
\label{sec:futurework}
    The following paragraphs outline the limitations we see with the current implementation and evaluation of \system, followed by what we consider meaningful avenues for future research.
    
\paragraph{Limitations} While we gained highly valuable insights with our user study, the effects of a time-constrained play session might not mirror everyday board game experiences.
        Furthermore, we consider the sample to be rather small and biased toward users with a higher affinity for technology.
Similarly, half of the participants can be considered experienced with digital fabrication, skewing the sample further.
        \system hands control over physical space to a hybrid board game, but did not enable full control and bi-directionality between digital and physical representations.
        For instance, customizing a figure did not have an effect on the digital avatar displayed on the game UI.
        As a system, \system is currently tied to a very specific hardware setup (especially the laser cutter) and a specific game.
        Making it generic for any and all hardware combinations, while similarly making it generically applicable to any type of game is likely to be a challenge.
        While some of these aspects limit the space we were able to explore with \system, we argue that it is an incredibly rich one, full of potential for hardware, software, and game-experience advances.

\paragraph{Future Work} \system can be improved from a \textbf{technical} perspective by adding additional sensing modules or fabrication capabilities~\cite{vasquezJubileeExtensibleMachine2020}.
The laser cutter could also be used for engraving onto players and the board, to create yet another tangible memento from a memorable game session.
            The narration and game master aspects of \system can also be enhanced through intelligence and adaptivity~\cite{triyasonExploringPotentialChatGPT2023,angFableRebornInvestigating2023}.
The \textbf{gameplay} of \system can also be improved further.
            In particular, we would like to further balance the game itself (or allow for balancing on-the-fly~\cite{rogersKickARExploringGame2018}).
            We further aim to facilitate more figure customization, which, ideally, is both precise and visually appealing, while also being done by the users themselves (i.e., not using pre-made designs as in many board games, e.g.,~\citegame{kingdomdeathmonster,warhammerthirdedition,gloomhaven,frosthaven}).
All aforementioned points are particularly relevant to enable longer scenarios or entire campaigns and increase the complexity of the game.
            We are further interested in \textbf{larger-scale studies} using \system. 
            In particular, the ratio of ``fabrication-experienced'' to ``fabrication-novice'' participants was not representative of a broader population~\cite{stemasovEnablingUbiquitousPersonal2021}.
            This is crucial to consider when treating \system as an implicit learning system, and not exclusively a game system~\cite{turakhiaReflectiveMakerUsing2022}.
            We currently do not see \system as an environment for explicit knowledge acquisition yet, but see a potential to embed and gamify this notion~\cite{lafreniereBlockstoCADCrossApplicationBridge2018, liGamiCADGamifiedTutorial2012}. 
            This would need more explicit aspects of feedback that may generalize to acquiring knowledge of digital fabrication processes~\cite{turakhiaReflectiveMakerUsing2022,kafaiHiLoTechGames2015,pearceEBeeMergingQuilting2016}.
            
 \section{Conclusion}
    \label{sec:conclusion}
We presented \system, a hybrid board game infused with and augmented through fabrication abilities to enable a tighter coupling between the physical and digital environment. 
    \system augments hybrid board games with the ability of machine-driven manipulation, destruction, and computational assessment of player-crafted artifacts. 
    This allows a hybrid board game to actively perceive and influence the physical play space, potentially leading to a stronger relationship between player and game experience, leveraging physical expression and creation and, similarly, the potential of physical loss and damage.

In a user study, we evaluated \system, confirming that it provides a highly engaging and enjoyable experience and that components like crafting, machine-driven motion, and potential destruction contribute positively to the overall experience.
    We further found that manually crafting equipment for one's figures may increase players' attachment to them, akin to personalization approaches in digital games.

    Our approach presents an alternative vision of how personal fabrication devices might become everyday appliances. 
    Instead of being general-purpose tools for solving mechanical problems, they become augmentations to existing concepts like board games, allowing them to mirror digital actions onto the physical environment and vice-versa.
    We consider approaches like \system to open up a rich and highly promising space for both (hybrid) board game design, and personal fabrication technologies alike.

\begin{acks}
    We thank our survey and study participants, the reviewers of the manuscript, and the contributors to the Printrun software ecosystem.
    We further thank Regan Mandryk for providing feedback on the concept.\newline
    \noindent This research was partially funded by the Deutsche Forschungsgemeinschaft (DFG, German Research Foundation) through the project \textit{``Democratic and Sustainable Personal Design and Fabrication through In-situ Co-Design and Previsualization''} (Project number: 525038300).\newline
    \noindent This work has been co-funded by the LOEWE initiative (Hesse, Germany) within the emergenCITY center.
\end{acks} 

\renewcommand{\bibnumfmt}[1]{[#1]}\bibliographystyle{ACM-Reference-Format}
\bibliography{zotero}

\renewcommand{\bibnumfmt}[1]{[\citegameprefix#1]}\bibliographystylegame{ACM-Reference-Format}
\bibliographygame{games}

\end{document}